\input epsf.hlp
\magnification=\magstep1
\overfullrule=0pt
\def\q#1{\lbrack #1 \rbrack}
\def\pano{\par\noindent}
\def\smno{\smallskip\noindent}
\def\meno{\medskip\noindent}
\def\bigno{\bigskip\noindent}
\def\o#1{\overline{#1}}
\def\pt{\partial}
\def\la{\langle}
\def\ra{\rangle}
\def\cF{{\cal{F}}}
\def\cO{{\cal{O}}}
\def\bz{\overline{z}}

\def\cl{\centerline}

\def\section#1{\leftline{\bf #1}\vskip-7pt\line{\hrulefill}}
\def\bibitem#1{\parindent=8mm\item{\hbox to 6 mm{$\q{#1}$\hfill}}}
\def\P{\,\hbox{\hbox to -0.2pt{\vrule height 6.5pt width .2pt\hss}\rm P}}
\def\BNT{\,\hbox{\hbox to -0.2pt{\vrule height 6.5pt width .2pt\hss}\rm N}}
\def\BRT{\,\hbox{\hbox to -0.2pt{\vrule height 6.5pt width .2pt\hss}\rm R}}
\def\BZT{{\rm Z{\hbox to 3pt{\hss\rm Z}}}}
\def\BZS{{\hbox{\sevenrm Z{\hbox to 2.3pt{\hss\sevenrm Z}}}}}
\def\BZSS{{\hbox{\fiverm Z{\hbox to 1.8pt{\hss\fiverm Z}}}}}
\def\BZ{{\mathchoice{\BZT}{\BZT}{\BZS}{\BZSS}}}
\def\BQT{\,\hbox{\hbox to -2.8pt{\vrule height 6.5pt width .2pt\hss}\rm Q}}
\def\BQS{\,\hbox{\hbox to -2.1pt{\vrule height 4.5pt width .2pt\hss}$
 \scriptstyle\rm Q$}}
\def\BQSS{\,\hbox{\hbox to -1.8pt{\vrule height 3pt width
 .2pt\hss}$\scriptscriptstyle \rm Q$}}

\def\BCT{\,\hbox{\hbox to -3pt{\vrule height 6.5pt width .2pt\hss}\rm C}}
\def\BCS{\,\hbox{\hbox to -2.2pt{\vrule height 4.5pt width .2pt\hss}$ 
 \scriptstyle\rm C$}}
\def\BCSS{\,\hbox{\hbox to -2pt{\vrule height 3.3pt width
 .2pt\hss}$\scriptscriptstyle \rm C$}}
\def\BC{{\mathchoice{\BCT}{\BCT}{\BCS}{\BCSS}}}
\def\bag{1}
\def\banks{2}
\def\berg{3}
\def\selfa{4}
\def\selfb{5}
\def\selfc{6}
\def\canda{7}
\def\candb{8}
\def\crem{9}
\def\dine{10}
\def\disgrea{11}
\def\disgreb{12}
\def\diska{13}
\def\diskb{14}
\def\diskc{15}
\def\diskd{16}
\def\gepe{17}
\def\greene{18}
\def\intri{19}
\def\kawit{20}
\def\kawai{21}
\def\sche{22}
\def\schz{23}
\def\schd{24}
\def\schv{25}
\def\silvera{26}
\def\silverb{27}
\def\wittena{28}
\def\wittenb{29}
\def\zamo{30}

\font\Large=cmr12 scaled \magstep3
\rm
\nopagenumbers
\pano
{\rightline {\vbox{\hbox{hep--th/9604140}
                   \hbox{BONN--TH--96--02}
                   \hbox{IFP--607--UNC}
                   \hbox{April 1996}}}}
\bigno\bigno
\centerline{\Large Exploring the Moduli Space of (0,2) Strings}
\vskip 1.2cm
\centerline{{Ralph Blumenhagen${}^1$}\ \ and \ \
            {Andreas Wi{\ss}kirchen${}^2$}}
\vskip 1.0cm
\centerline{${}^1$ \it Institute of Field Physics, Department of Physics
and Astronomy,}
\centerline{\it University of North Carolina,
Chapel Hill NC 27599--3255, USA}
\vskip 0.1cm
\centerline{${}^2$ \it Physikalisches Institut der Universit\"at Bonn,
Nu{\ss}allee 12, 53115 Bonn, Germany}
\vskip 1.0cm
\centerline{\bf Abstract}
\meno
We use an exactly solvable $(0,2)$ supersymmetric conformal field theory
with gauge group $SO(10)$ to investigate the superpotential of the 
corresponding classical string vacuum. We provide evidence that the 
rational point lies in the Landau--Ginzburg phase of the linear 
$\sigma-$model and calculate exactly all three-- and four--point functions
of the gauge singlets. These couplings already put severe constraints on 
the possible flat directions of the superpotential. Finally, we contemplate
about the flat direction related to K\"ahler deformations of the underlying
linear $\sigma-$model.
\footnote{}
{\pano
${}^1$ e--mail:\ blumenha@physics.unc.edu
\pano
${}^2$ e--mail:\ wisskirc@avzw02.physik.uni--bonn.de}
\vfill\eject
\footline{\hss\tenrm\folio\hss}
\pageno=1
\pano
\section{1.\ Introduction}
\meno
Mainly during the last three years progress has been made in showing that 
the class of $(2,2)$ supersymmetric string compactifications is only
a small subset of all four--dimensional perturbative heterotic string vacua
featuring $N=1$ space--time supersymmetry 
[\banks--\selfb,\disgrea,\diska--\diskc,\kawai,\silvera,\silverb,\wittenb].
It was long believed that the general class of $(0,2)$ strings might not 
be solutions of the string equations of motion at all, for these models 
could receive destabilizing instanton corrections [\dine,\wittena].
However, in a paper by E.\ Silverstein and E.\ Witten [\silverb] it was 
argued that for the class of linear $\sigma-$models such terms in the 
superpotential cannot occur due to the absence of singularities in the
singlet couplings. Shortly afterwards in [\selfa] we constructed
heterotic exactly solvable $(0,2)$ superconformal field theories (SCFTs)
exhibiting all the properties required for $(0,2)$ string vacua. For
instance, the phenomenologically most interesting feature is that the
gauge group is not restricted to $E_6$ as in the $(2,2)$ case [\canda]
but can also be $SO(10)$ or $SU(5)$ [\wittena]. The main difficulty
turned out to really identify SCFTs with special points in the moduli
space of a Calabi--Yau $\sigma-$model with a choice of a stable,
holomorphic vector bundle for the left moving $\sigma-$model fermions.
In [\selfb] for at least three $N=1$ models such an identification has
been shown to be possible, including the $SO(10)$ model we will focus on
in this paper. Moreover, for the class of $N=2$ space--time
supersymmetric strings it was furthermore possible to identify all
constructed SCFTs with certain bundles on $K_3\times T_2$ [\selfc]. 
\pano 
In this paper we will investigate the moduli space of a concrete $(0,2)$ 
model by for the first time calculating all three-- and four--point
functions of the gauge singlets at the exactly solvable point. In [\diskb]
using the Landau--Ginzburg description of the quintic in $\BC{\rm P}[4]$
it was already shown that besides the well known $(2,2)$ moduli space
[\candb] containing the complex and K\"ahler deformations there are flat 
directions in the superpotential due to elements of $H^1({\rm End}(T))$,
as well. At least all states coming from untwisted sectors of the LG 
orbifold are mutually integrable. In [\silverb] it was argued, and 
exemplified again for the case of the quintic, that even more is true.
All deformations related to parameters of the linear $\sigma-$model are
moduli for any value of the K\"ahler class. At the Landau--Ginzburg point
this includes some moduli from twisted sectors. Thus, for the quintic in
$\BC{\rm P}[4]$ there is a 326--dimensional $(0,2)$ moduli space
containing a 102--dimensional $(2,2)$ subspace.
\pano
One advantage of the knowledge of an exactly solvable model for a $(0,2)$ 
string vacuum is that it 
allows one to make definite statements about nonvanishing superpotential
couplings, in general at least to finite order in the superpotential.
We study in detail the $(0,2)$ model with gauge group $SO(10)$ which was 
first constructed in [\selfa] and then related to the Calabi--Yau manifold
$\P_{1,1,1,1,2,2}[4\ 4]$ with the bundle $V(1,1,1,1,1;5)$ in [\selfb].
After identifying our SCFT with a special point in the Landau--Ginzburg 
sector, we calculate the couplings of all singlets to fourth order and 
find that they do not all vanish. At the Landau--Ginzburg point there occur
more than the 329 massless singlets expected from the large radius limit.
Thus, not all singlets at the Landau--Ginzburg point correspond to
flat directions of the space--time superpotential. Unlike the $(2,2)$ 
case [\disgreb,\gepe], at small radius there exists no algebraic 
distinction among the complex, K\"ahler and bundle moduli. However, in 
our special model, requiring certain properties expected for the 
K\"ahler modulus and using F--flatness and D--flatness for the special 
enhanced gauge symmetry, at least to lowest order the possible
candidates for the deformation of the radius are highly restricted.  
\pano
This paper is organized as follows. In section 2.\ we review the 
construction of the $(0,2)$ SCFT. In section 3.\ we give evidence that 
the SCFT lies in the Landau--Ginzburg phase of the corresponding linear
$\sigma-$model. In section 4.\ we make use of the SCFT to calculate all
holomorphic three-- and four--point functions of the $SO(10)$ singlets.
Finally, in section 5.\ we use these results to restrict the form of the
K\"ahler modulus and to speculate about new flat directions in the 
superpotential not belonging to the moduli space of the linear 
$\sigma-$model.
\meno
\section{2.\ The exactly solvable model}
\smno
To begin with, we briefly review the model we will deal with in this 
paper. The details of the construction of general $(0,2)$ SCFTs can be 
found in [\selfa,\selfb]. In order to achieve heterotic modular invariant
partition functions we made use of the technique of simple currents 
[\intri,\sche--\schv]. In light cone gauge the starting point
of the construction is the tensor product of CFTs as shown in Table 2.1.
\smno
\cl{\vbox{
\hbox{\vbox{\offinterlineskip
\def\tablespace{height2pt&\omit&&\omit&&\omit&\cr}
\def\tablerule{\tablespace\noalign{\hrule}\tablespace}
\hrule\halign{&\vrule#&\strut\hskip0.2cm\hfil#\hfill\hskip0.2cm\cr
\tablespace
& part && $c$ && $\o{c}$ &\cr
\tablerule\tablerule
& $4D$ space--time, $X^{\mu}$ && $2$ && $2$ & \cr
\tablerule
& $N=2$ Virasoro && $9$ && $9$ &\cr
\tablerule
& $U(1)_2$ && $1$ && $1$ &\cr
\tablerule
& gauge group\ $SO(8)\times E_8$ && $12$ && $12$ &\cr
\tablespace}\hrule}}
\hbox{\hskip 0.5cm Table 2.1 \hskip 0.5cm Underlying CFT for $SO(10)$}}}
\smno
Introducing a special set of simple currents guaranteeing
all properties we want to have for $(0,2)$ models like left
moving $N=2$ world sheet supersymmetry, projection onto
even $U(1)$ charges and an extension of the gauge group from
$SO(8)$ to $SO(10)$, we obtain modular invariant partition
functions of the following form:
$$ Z\sim\vec{\chi}(\tau)\,M(J_{GSO_l})\,\prod_j M(\Upsilon_j)\,\,
  M(J_{GSO_r})\,\prod_i M(J_i)\,\,M(J_{(SO(8)\to SO(10))})\,
  \vec{\chi}(\o\tau).\eqno(2.1)$$
In order to really get $(0,2)$ models one has to choose some
model dependent simple currents $\Upsilon_j$, which prevent
all the symmetries implemented on the right moving side
from acting also on the left moving side. In the model discussed
in [\selfb], the internal $N=2$ part consists of five copies of the 
$k=3$ minimal $N=2$ model (N2Vir(k=3)) and $\Upsilon$ is chosen to be
$$ \Upsilon=\Phi^3_{0,-1}\otimes\left(\Phi^0_{0,0}\right)^4
 \otimes\Phi^{U(1)_2}_{1,2}\otimes\Phi^{SO(8)}_0,\eqno(2.2) $$
where $\Phi^l_{m,s}$ denotes the highest weight representations of the 
$k=3$ minimal model. 
Thus $\Upsilon$ acts nontrivially only on the first factor of the 
internal tensor product and on the $U(1)_2$ part. The massless 
spectrum contains the usual $N=1$ supergravity multiplet, chiral
multiplets in the four possible representations of $SO(10)_1$
and also some vector multiplets. There are precisely $N_{16}=80$
chiral fields in the spinor representation, $N_{10}=74$ chiral
fields in the vector representation and $N_{1}=350$ 
chiral fields in the singlet representation of $SO(10)$. 
Furthermore, besides the gauge bosons of $SO(10)\times E_8$ the
spectrum contains $N_g=7$ further vector multiplets. Since the
simple current $\Upsilon$ itself and its charge conjugate 
are two of these seven further gauge bosons, the special
enhanced gauge group cannot simply be $U(1)^7$. 
It is easy to see that the three fields of conformal dimension one
$$\eqalignno{J^\pm(z) &= \Phi^3_{\mp 1}(z) \otimes 
  e^{\mp i \sqrt{5\over 3} \phi_1(z)}\otimes
  e^{\pm i {1\over 2} \phi_{U(1)_2}(z)}, &(2.3)\cr
  J^3(z) &= {1\over 2}\left( -5\ j_1(z) + 3\ j_{U(1)_2}(z) \right) &\cr }$$
satisfy the $SU(2)$ Kac--Moody algebra at level $k=3$. $\Phi^3_{\mp 1}(z)$ 
are primary fields of the $k=3$ parafermionic model and
the $U(1)$ currents of the first N2Vir(k=3) model and the $U(1)_2$ model
are $j_1=i\sqrt{3\over 5} \pt\phi_1$ and $j_{U(1)_2}=i \pt\phi_{U(1)_2}$,
respectively. Thus, the complete gauge group of this model is 
$G=SO(10)\times E_8\times SU(2)_3\times U(1)^4$ and the massless spectrum 
should also fit into the four allowed representations of $SU(2)_3$.
\pano
In [\selfb] we have listed the explicit form of the massless states in the
spinor and vector representation of $SO(10)$ and have given a monomial
representation of these states such that the Yukawa couplings 
$\la 10\ 16\ 16\ra$ could be written as a monomial ring 
$ \BC [x_i,y_j]/I$. The $x_i$ are four coordinates of weight one and the
$y_j$ are two coordinates of weight two. The ideal $I$ is generated by the 
relations $x_i^4=0$ and $y_j\, y_k=0$. The same monomial
ring appears as the cohomological ring of the Calabi--Yau manifold
$\P_{1,1,1,1,2,2}[4\ 4]$ with the gauge bundle $V(1,1,1,1,1;5)$.
Thus, we concluded that the exactly solvable SCFT describes a certain point
in the moduli space of this $(0,2)$ model. 
However, we have not determined to which point the SCFT corresponds, namely
what the form is of the quasi--homogeneous polynomials $W_{1,2}(x_i,y_j)$
and $F_{1,\ldots,5}(x_i,y_j)$ defining the complete intersection CY and
the vector bundle $V$, respectively. Furthermore, one has to know at which
radius $r$ of the K\"ahler modulus the model lives. In order to answer 
these questions we also have to take the singlet fields into account. 
\pano
A singlet in the (--1) ghost picture is of the general form 
$$ V_{-1}(z,\bz)=e^{-\rho(\bz)}\cO_1(z,\bz)\ 
 \cF(z)\ e^{ikX(z,\bz)},\eqno(2.4)$$
where $\cO_1$ is an internal operator of the $N=2$ theory with 
$(c,\o c)=(9,9)$ and $\cF$ denotes the left moving $U(1)_2$ part. The 
product $\cO_{1}\,\cF$ has overall conformal dimension 
$(h,\o{h})=(1,{1\over2})$ and charge $(q,\o{q})=(0,-1)$. In Table 2.2 and
2.3 we list the explicit form of these internal operators for all the 350
singlets occurring in the model. The degeneracy is due to three reasons.
Firstly, there is the $S_4$ permutation symmetry of the last four 
N2Vir(k=3) tensor factors. Secondly, we have the four allowed $SU(2)_3$
representations with degenerated ground states of dimension one to four.
In Table 2.2 and 2.3 we always list the state with highest value of the
$U(1)_2$ quantum number. Thus, the other states in the $SU(2)$ multiplet
can be obtained by applying successively $J^-$. Finally, whenever a state
like, for instance\footnote{$^1$}{We use the notation introduced in 
[\selfb].\ $\left[l~\matrix{m&s\cr \o{m}&\o{s}\cr}\right]$ denotes a 
primary field of N2Vir(k=3) and $[m]$ denotes one of the four primary 
fields of $U(1)_2$.}
$$ \left[0~\matrix{0&0\cr 0&0\cr}\right]
            \left[3~\matrix{3&2\cr 3&0\cr}\right]
            \left[2~\matrix{2&0\cr 2&0\cr}\right]
            \left[0~\matrix{0&0\cr 0&0\cr}\right]
            \left[0~\matrix{0&0\cr 0&0\cr}\right]
            [0].\eqno(2.5)$$
occurs there is also a state
$$ \left[0~\matrix{0&0\cr 0&0\cr}\right]
            \left[3~\matrix{3&0\cr 3&0\cr}\right]
            \left[2~\matrix{2&2\cr 2&0\cr}\right]
            \left[0~\matrix{0&0\cr 0&0\cr}\right]
            \left[0~\matrix{0&0\cr 0&0\cr}\right]
            [0]. \eqno(2.6)$$
The left moving $G^i=\left[0~\matrix{0&2\cr 0&0\cr}\right]$ can be 
permuted among all nonzero fields in the last four $N=2$ tensor factors. 
In accordance to [\diskb] we denote the untwisted fields by $S$ and the
twisted ones by $S'$.
\meno
\cl{\vbox{
\hbox{\vbox{\offinterlineskip
\def\tablespace{height2pt&\omit&&\omit&&\omit&&\omit&\cr}
\def\tablerule{\tablespace\noalign{\hrule}\tablespace}
\hrule\halign{&\vrule#&\strut\hskip0.2cm\hfil#\hfill\hskip0.2cm\cr
\tablespace
&Type&&\hskip3.7cm$\cO_{1}$\hskip3.5cm$\cF$&&$SU(2)$ rep.&&deg.&\cr
\tablerule\tablerule
& $S_a$ && $\left[0~\matrix{0&0\cr 0&0\cr}\right]
            \left[3~\matrix{3&2\cr 3&0\cr}\right]
            \left[2~\matrix{2&0\cr 2&0\cr}\right]
            \left[0~\matrix{0&0\cr 0&0\cr}\right]
            \left[0~\matrix{0&0\cr 0&0\cr}\right]
            [0]$
&& $1$ && $24$ & \cr
\tablerule 
& $S_b$ && $\left[0~\matrix{0&0\cr 0&0\cr}\right]
            \left[3~\matrix{3&2\cr 3&0\cr}\right]
            \left[1~\matrix{1&0\cr 1&0\cr}\right]
            \left[1~\matrix{1&0\cr 1&0\cr}\right]
            \left[0~\matrix{0&0\cr 0&0\cr}\right]
            [0]$
&& $1$ && $36$ & \cr
\tablerule 
& $S_c$ && $\left[0~\matrix{0&0\cr 0&0\cr}\right]
            \left[2~\matrix{2&2\cr 2&0\cr}\right]
            \left[2~\matrix{2&0\cr 2&0\cr}\right]
            \left[1~\matrix{1&0\cr 1&0\cr}\right]
            \left[0~\matrix{0&0\cr 0&0\cr}\right]
            [0]$
&& $1$ && $36$ & \cr
\tablerule 
& $S_d$ && $\left[0~\matrix{0&0\cr 0&0\cr}\right]
            \left[2~\matrix{2&2\cr 2&0\cr}\right]
            \left[1~\matrix{1&0\cr 1&0\cr}\right]
            \left[1~\matrix{1&0\cr 1&0\cr}\right]
            \left[1~\matrix{1&0\cr 1&0\cr}\right]
            [0]$
&& $1$ && $16$ & \cr
\tablerule 
& $S_e$ && $\left[3~\matrix{3&0\cr 3&0\cr}\right]
            \left[2~\matrix{2&0\cr 2&0\cr}\right]
            \left[0~\matrix{0&0\cr 0&0\cr}\right]
            \left[0~\matrix{0&0\cr 0&0\cr}\right]
            \left[0~\matrix{0&0\cr 0&0\cr}\right]
            [2]$
&& $4$ && $16$ & \cr
\tablerule 
& $S_f$ && $\left[3~\matrix{3&0\cr 3&0\cr}\right]
            \left[1~\matrix{1&0\cr 1&0\cr}\right]
            \left[1~\matrix{1&0\cr 1&0\cr}\right]
            \left[0~\matrix{0&0\cr 0&0\cr}\right]
            \left[0~\matrix{0&0\cr 0&0\cr}\right]
            [2]$
&& $4$ && $24$ & \cr
\tablerule 
& $S_g$ && $\left[2~\matrix{2&0\cr 2&0\cr}\right]
            \left[3~\matrix{3&2\cr 3&0\cr}\right]
            \left[0~\matrix{0&0\cr 0&0\cr}\right]
            \left[0~\matrix{0&0\cr 0&0\cr}\right]
            \left[0~\matrix{0&0\cr 0&0\cr}\right]
            [0]$
&& $2$ && $8$ & \cr
\tablerule 
& $S_h$ && $\left[2~\matrix{2&0\cr 2&0\cr}\right]
            \left[2~\matrix{2&2\cr 2&0\cr}\right]
            \left[1~\matrix{1&0\cr 1&0\cr}\right]
            \left[0~\matrix{0&0\cr 0&0\cr}\right]
            \left[0~\matrix{0&0\cr 0&0\cr}\right]
            [0]$
&& $2$ && $48$ & \cr
\tablerule 
& $S_i$ && $\left[2~\matrix{2&0\cr 2&0\cr}\right]
            \left[1~\matrix{1&2\cr 1&0\cr}\right]
            \left[1~\matrix{1&0\cr 1&0\cr}\right]
            \left[1~\matrix{1&0\cr 1&0\cr}\right]
            \left[0~\matrix{0&0\cr 0&0\cr}\right]
            [0]$
&& $2$ && $24$ & \cr
\tablerule 
& $S_j$ && $\left[1~\matrix{1&0\cr 1&0\cr}\right]
            \left[3~\matrix{3&0\cr 3&0\cr}\right]
            \left[1~\matrix{1&0\cr 1&0\cr}\right]
            \left[0~\matrix{0&0\cr 0&0\cr}\right]
            \left[0~\matrix{0&0\cr 0&0\cr}\right]
            [2]$
&& $3$ && $36$ & \cr
\tablerule 
& $S_k$ && $\left[1~\matrix{1&0\cr 1&0\cr}\right]
            \left[2~\matrix{2&0\cr 2&0\cr}\right]
            \left[2~\matrix{2&0\cr 2&0\cr}\right]
            \left[0~\matrix{0&0\cr 0&0\cr}\right]
            \left[0~\matrix{0&0\cr 0&0\cr}\right]
            [2]$
&& $3$ && $18$ & \cr
\tablerule 
& $S_l$ && $\left[1~\matrix{1&0\cr 1&0\cr}\right]
            \left[2~\matrix{2&0\cr 2&0\cr}\right]
            \left[1~\matrix{1&0\cr 1&0\cr}\right]
            \left[1~\matrix{1&0\cr 1&0\cr}\right]
            \left[0~\matrix{0&0\cr 0&0\cr}\right]
            [2]$
&& $3$ && $36$ & \cr
\tablerule 
& $S_m$ && $\left[1~\matrix{1&0\cr 1&0\cr}\right]
            \left[1~\matrix{1&0\cr 1&0\cr}\right]
            \left[1~\matrix{1&0\cr 1&0\cr}\right]
            \left[1~\matrix{1&0\cr 1&0\cr}\right]
            \left[1~\matrix{1&0\cr 1&0\cr}\right]
            [2]$
&& $3$ && $3$ & \cr
\tablespace}\hrule}}
\hbox{\hskip 0.5cm Table 2.2 \hskip 0.5cm Untwisted Singlets}}}
\meno
These are exactly 325 of the 350 singlets. The remaining 25 states occur
in twisted sectors of the $GSO_r$ projection. 
\meno
\cl{\vbox{
\hbox{\vbox{\offinterlineskip
\def\tablespace{height2pt&\omit&&\omit&&\omit&&\omit&\cr}
\def\tablerule{\tablespace\noalign{\hrule}\tablespace}
\hrule\halign{&\vrule#&\strut\hskip0.2cm\hfil#\hfill\hskip0.2cm\cr
\tablespace
&Type&&\hskip4.5cm$\cO_{1}$\hskip4.5cm$\cF$&&rep.&&deg.&\cr
\tablerule\tablerule
& $T'$ && $\left[0~\matrix{1&1\cr 0&0\cr}\right]
            \left[2~\matrix{-2&0\cr 2&0\cr}\right]
            \left[1~\matrix{-3&-2\cr 1&0\cr}\right]
            \left[1~\matrix{-3&-2\cr 1&0\cr}\right]
            \left[1~\matrix{-3&-2\cr 1&0\cr}\right]
            [1]$
&& $1$ && $4$ & \cr
\tablerule
& $S'_a$ && $\left[3~\matrix{-4 &-1\cr 3&0\cr}\right]
            \left[1~\matrix{-1&0\cr 1&0\cr}\right]
            \left[1~\matrix{-1&0\cr 1&0\cr}\right]
            \left[0~\matrix{-2&-2\cr 0&0\cr}\right]
            \left[0~\matrix{-2&-2\cr 0&0\cr}\right]
            [1]$
&& $1$ && $6$ & \cr
\tablerule
& $S'_b$ && $\left[2~\matrix{5 & 5\cr 2&0\cr}\right]
            \left[1~\matrix{-1&0\cr 1&0\cr}\right]
            \left[1~\matrix{-1&0\cr 1&0\cr}\right]
            \left[1~\matrix{-1&0\cr 1&0\cr}\right]
            \left[0~\matrix{-2&-2\cr 0&0\cr}\right]
            [1]$
&& $3$ && $12$ & \cr
\tablerule
& $S'_c$ && $\left[1~\matrix{-1 & -2\cr 1&0\cr}\right]
            \left[1~\matrix{-1&0\cr 1&0\cr}\right]
            \left[1~\matrix{-1&0\cr 1&0\cr}\right]
            \left[1~\matrix{-1&0\cr 1&0\cr}\right]
            \left[1~\matrix{-1&0\cr 1&0\cr}\right]
            [0]$
&& $3$ && $3$ & \cr
\tablespace}\hrule}}
\hbox{\hskip 0.5cm Table 2.3 \hskip 0.5cm Twisted Singlets}}}
\meno
These numbers of untwisted and twisted singlets have to be compared with 
the numbers of singlets in the LG phase of the corresponding linear 
$\sigma-$model. It has been shown in [\diska] that for a generic choice 
of the $W_i(x,y)$ and $F_i(x,y)$ the model contains 339 singlets, 
318 of which are untwisted. Note that in the Calabi--Yau limit the
number of singlets is only 
$$H^1(M,T)+H^1(M,T^*)+H^1(M,{\rm End}(V))=73+1+255=329.\eqno(2.7)$$
\meno
\section{3.\ The Landau--Ginzburg phase}
\meno
In order to identify the SCFT with the Landau--Ginzburg phase of the
linear $\sigma-$model we have at least to show that the massless
spectra are the same. For our model of interest the $(0,2)$ superpotential
is
$$ S_{\cal W} =\int d^2 z d\theta\ \left( \Sigma_j W_j(X,Y) 
               + \Lambda_a F_a(X,Y)\, \right), \eqno(3.1)$$
where $\Sigma_{1,2}$ and $\Lambda_{1,\ldots,5}$ are Fermi superfields and 
$X_{1,\ldots,4}$ and $Y_{1,2}$ are chiral superfields. $ W_{1,2}(X,Y)$ 
and $F_{1,\ldots,5}(X,Y)$ are quasihomogenous polynomials of degree four.
In the Landau--Ginzburg phase there exists a right $U(1)$ R--symmetry with 
charges $\o{q}$ and a left $U(1)$ symmetry with charges $q$. In Table 3.1
we list all the charges of the fields involved in the calculation of the 
massless spectrum.
\meno
\cl{\vbox{\hbox{
\vbox{
\hbox{\vbox{\offinterlineskip
\def\tablespace{height2pt&\omit&&\omit&&\omit&\cr}
\def\tablerule{\tablespace\noalign{\hrule}\tablespace}
\hrule\halign{&\vrule#&\strut\hskip0.2cm\hfil#\hfill\hskip0.2cm\cr
\tablespace
& field && $q$ && $\o{q}$ &\cr
\tablerule\tablerule
& $x_{1,\ldots,4}$ && ${1\over 5}$ && ${1\over 5}$ & \cr
\tablerule
& $y_{1,2}$ && ${2\over 5}$ && ${2\over 5}$ &\cr
\tablerule
& $\sigma_{1,2}$ && $-{4\over 5}$ && ${1\over 5}$ &\cr
\tablerule
& $\lambda_{1,\ldots,5}$ && $-{4\over 5}$ && ${1\over 5}$ &\cr
\tablespace}\hrule}} }
\hskip 1cm
\vbox{
\hbox{\vbox{\offinterlineskip
\def\tablespace{height2pt&\omit&&\omit&&\omit&\cr}
\def\tablerule{\tablespace\noalign{\hrule}\tablespace}
\hrule\halign{&\vrule#&\strut\hskip0.2cm\hfil#\hfill\hskip0.2cm\cr
\tablespace
& field && $q$ && $\o{q}$ &\cr
\tablerule\tablerule
& $\o x_{1,\ldots,4}$ && $-{1\over 5}$ && $-{1\over 5}$ & \cr
\tablerule
& $\o y_{1,2}$ && $-{2\over 5}$ && $-{2\over 5}$ &\cr
\tablerule
& $\o\sigma_{1,2}$ && ${4\over 5}$ && $-{1\over 5}$ &\cr
\tablerule
& $\o\lambda_{1,\ldots,5}$ && ${4\over 5}$ && $-{1\over 5}$ &\cr
\tablespace}\hrule}}} }
\hbox{\hskip 0.5cm Table 3.1 \hskip 0.5cm Left and right charges } }}
\meno
For generic choices of the polynomials the massless spectrum has been 
calculated for the Landau--Ginzburg orbifold in [\diska]. There are 80 
chiral superfields in the spinor representation, 74 superfields in the 
vector representation and 339 superfields in the singlet representation 
of SO(10). The question is, whether there exists a choice of the 
polynomials $W_j$ and $F_a$ such that there occur 350 singlets, 325 
untwisted and 25 twisted, accompanied by an enhanced gauge group of 
dimension seven. In [\selfb] we have already made the following guess 
for the form of the constraints:
$$\eqalignno{W_1(X,Y)&=\sum_{i=1}^4X_i^4\ +\ \sum_{j=1}^2 Y_j^2,
 \quad\quad\ \ W_2(X,Y)=\sum_{i=1}^4 i\,X_i^4\ +\ \sum_{j=1}^2 j\ Y_j^2 
 &\cr F_i(X,Y)&=X_i^4\quad{\rm for}\ i\in\lbrace 1,\ldots,4\rbrace,
 \quad F_5(X,Y) = Y_1\,Y_2, &(3.2)\cr }$$
which was motivated by the fact that the exactly solvable model has the 
permutation symmetry $S_4$ in the last four $N=2$ tensor factors. In the 
Landau--Ginzburg model the $U(1)$ gauge symmetry of the linear 
$\sigma-$model is spontaneously broken to a finite group $\BZ_5$, so that
one actually deals with an orbifold theory. Furthermore, to get a 
heterotic string theory one has to combine the internal Landau--Ginzburg 
sector with the linear part of the gauge group, which is $SO(8)$ in our 
case. The GSO projection then selects states with $g=1$ for
$$ g={\rm exp}(-i\pi J_0)\times (-1)^\lambda. \eqno(3.3)$$
$J_0$ is the left $U(1)$ charge and $(-1)^\lambda$ denotes the charges of
the different $SO(8)$ representations. The resulting orbifold has sectors
twisted by $g^k$ for $k=0,\ldots,9$. If $k$ is even we call them $(R,R)$ 
sectors and if $k$ is odd we call them $(NS,R)$ sector. Finally, since one
is only interested in the massless sector of the string model, one can 
employ a Born--Oppenheimer approximation and truncate the fields to their
lowest excited modes. The right moving $N=2$ algebra 
$$\lbrace\o{Q}_-,\o{Q}_+\rbrace=\o{L}_0,\quad\o{Q}_-^2=\o{Q}_+^2=0
 \eqno(3.4)$$
tells us that massless states are given by the cohomology of $\o{Q}_+$. 
There is an expression even off--criticality for this operator in terms 
of the fundamental fields in the Lagrangian:
$$\o{Q}_+=i\int(i\o\psi^i\pt\phi_i+{\cal W}|_{\theta=0}).\eqno(3.5)$$
By splitting this into $\o{Q}_+=\o{Q}_{+,r} + \o{Q}_{+,l}$ it was shown in 
[\kawit] that one can simply calculate the cohomology of $\o{Q}_{+,l}$ in
the cohomology of $\o{Q}_{+,r}$. In order to get more insight into these 
methods the interested reader may take a look into [\diska,\kawit]. By 
going through the calculation of the massless spectrum carried out in
[\diska], one realizes that there are really more states for the choice 
of the polynomials in (3.2). Table 8 of [\diska] shows that the massless 
singlets get modified in the way described in Table 3.2.
\meno
\cl{\vbox{
\hbox{\vbox{\offinterlineskip
\def\tablespace{height2pt&\omit&&\omit&&\omit&\cr}
\def\tablerule{\tablespace\noalign{\hrule}\tablespace}
\hrule\halign{&\vrule#&\strut\hskip0.2cm\hfil#\hfill\hskip0.2cm\cr
\tablespace
& $SO(8)\times U(1)$ && $k$ && State &\cr
\tablerule\tablerule
& $1_0$ && $1$ && $P_4(\Phi^i_{-{q_i\over 2}})\lambda^a_{-{3\over 5}}|0\ra,
         \ P_4(\Phi^i_{-{q_i\over 2}})\sigma^j_{-{3\over 5}}|0\ra$ & \cr
\tablerule
& $1_0$ && $3$ && $\lambda^a_{-{1\over 5}}\lambda^b_{-{1\over 5}}|0\ra,\
                   \lambda^a_{-{1\over 5}}\sigma^j_{-{1\over 5}}|0\ra,\
                   \sigma^1_{-{1\over 5}}\sigma^2_{-{1\over 5}}|0\ra$ &\cr
\tablerule
& $1_0$ && $5$ && $\o\lambda^a_{0}\,\o\lambda^b_{0}\,\o\lambda^c_{0}|0\ra\ 
  {\rm for}\ a,b,c\in \lbrace 1,\ldots,4\rbrace$ &\cr
\tablespace}\hrule}}
\hbox{\hskip 0.5cm Table 3.2 \hskip 0.5cm Massless LG singlets}}}
\meno
$P_4(\Phi^i_{-{q_i\over 2}})$ denotes a polynomial of weight four in the 
six coordinates $x_i, y_j$. The untwisted states for $k=1$ have to be 
considered modulo the equivalence relations
$$\eqalignno{ &F_a(\Phi^i_{-{q_i\over 2}})\lambda^a_{-{3\over 5}}|0\ra \sim
   F_a(\Phi^i_{-{q_i\over 2}})\sigma^j_{-{3\over 5}}|0\ra \sim 0 & \cr
   &W_j(\Phi^i_{-{q_i\over 2}})\lambda^a_{-{3\over 5}}|0\ra \sim
   W_j(\Phi^i_{-{q_i\over 2}})\sigma^j_{-{3\over 5}}|0\ra \sim 0 & \cr
   &P_2(\Phi^i_{-{q_i\over 2}})\left[ {\pt F_a\over \pt 
   y^{1,2}_{-{1\over 5}}} \lambda^a_{-{3\over 5}} + {\pt W_j\over \pt 
   y^{1,2}_{-{1\over 5}} } \sigma^j_{-{3\over 5}} \right]|0\ra \sim 0 
   &(3.6)\cr &P_1(\Phi^i_{-{q_i\over 2}})\left[ {\pt F_a\over \pt 
   x^{1,\ldots,4}_{-{1\over 10}} } \lambda^a_{-{3\over 5}} + 
   {\pt W_j\over \pt x^{1,\dots,4}_{-{1\over 10}} }
   \sigma^j_{-{3\over 5}} \right]|0\ra \sim 0.&\cr }$$
For generic $W_j$ and $F_a$ there are 318 such states. However, for the 
symmetric choice in (3.2) we obtain 325 states, which are inevitably 
accompanied by seven further gauge fields. Furthermore, there occur four 
singlet fields from the $k=5$ sector, which are not present at a generic 
point in the moduli space. Thus, together with the 21 states from the
$k=3$ twisted sector there are 25 twisted singlets, the same number as for
the exactly solvable model. This brief excursion to $(0,2)$ 
Landau--Ginzburg models has provided some more evidence that we can 
identify the exactly solvable model with a $(0,2)$ Landau--Ginzburg model
naturally appearing in the $r\to-\infty$ limit of a linear $\sigma-$model. 
In the following section we will calculate all three-- and four--point 
functions of the SCFT.
\meno
\section{4.\ Singlet couplings in the superpotential}
\meno
Unlike, for instance, the quintic and the corresponding Landau--Ginzburg
model, in our $(0,2)$ case the number of $SO(10)$ singlets in the 
large radius limit is different from the number of singlets in the
Landau--Ginzburg phase. Since we are now also equipped with an explicit
SCFT description, it is possible to investigate moduli
in the neighbourhood of the exactly solvable point. 
Thus, we are looking for integrable marginal deformations of the SCFT
preserving the right moving $N=2$ world sheet supersymmetry. 
From the space--time point of view this is equivalent to searching for
flat directions in the four--dimensional effective low energy $N=1$ 
space--time supersymmetric field theory. The scalar potential for such 
supergravity theories is generally known as [\bag,\crem]
$$U=e^{{\cal K}}\left(D^iW\,G_{ij^{*}}^{-1}\,D^{j^*}W^*-3\,W\right)+
      {1\over 2}\sum_a(D^a)^2,\eqno(4.1)$$
where ${\cal K}(\phi_i,\phi^*_i)$ is the K\"ahler potential, 
$W(\phi_i)$ the holomorphic superpotential and $T^a$ generators of the 
gauge group. The covariant derivative is given by
$$D^iW={\pt W\over\pt\phi_i}+{\pt{\cal K}\over\pt\phi_i}W\eqno(4.2)$$ 
and $D^a$ are auxiliary fields in the vector multiplets of the gauge bosons
$$ D^a=-{g_a\over 2}\,\left({\pt{\cal K}\over\pt\phi_i}\,T^a\phi_i+\phi^*_i
 \,T^a\,{\pt{\cal K}\over\pt\phi^*_i}\right).\eqno(4.3)$$
In the lowest order, the renormalizable field theory limit, ${\cal K}$
is flat and the scalar potential takes the form
$$ U=\sum_a {g_a^2\over 2} |D^a|^2 + \sum_i |F_i|^2= \sum_a 
         {g_a^2\over 2}\, (\phi^*_i T^a \phi_i)^2
     + \sum_i \left| {\pt W\over \pt\phi_i} \right|^2 \eqno(4.4) $$
where $W$ contains only cubic couplings. In order for the scalar potential
to vanish both the D--terms and the F--terms have to be zero.
\pano
For gauge singlets the condition of D--flatness is satisfied 
automatically and one has only to check F--flatness for the superpotential.
However, even though we are dealing mostly with $SO(10)$ singlets
we have to cope with the D--terms arising from the enhanced
gauge symmetry $SU(2)\times U(1)^4$. 
The superpotential of the low energy effective field theory can
be determined in the SCFT by calculating correlation functions of the 
corresponding vertex operators on the $S^2$ world sheet. 
For the following discussion one has to have in mind that the 
vertex operators in the SCFT geometrically are tangent vectors
along the moduli space. Suppose one finds a set of scalars 
$\lbrace S_i\rbrace$, such that all D--terms vanish and all 
superpotential couplings of the form $F(S)$ and $F(S)S'$ are zero for 
all scalars $S'$ in the model. Then the entire set $\lbrace S_i\rbrace$
are flat directions and define bona fide moduli of the theory. 
However, if one has a set of scalars for which not all terms in 
the scalar potential vanish, then one has to be very careful in
drawing any conclusions. It does not necessarily mean that there are
no flat directions at all. This can be seen by studying the following 
well known example:\ Consider the simple case of a complex boson
with a global $U(1)$ symmetry and the Higgs potential
$$V(\phi)=\lambda\left(|\phi|^2-{m^2\over2\lambda}\right)^2.\eqno(4.5)$$ 
Now, expanding around one minimum $\la\phi\ra=v=\sqrt{m^2\over 2 \lambda}$
in the usual way
$$ \phi=\eta+v +i \chi \eqno(4.6)$$
one finds in the potential 
$$V(\phi)=\lambda\left(4v^2\eta^2+4v\eta(\eta^2+\chi^2)+(\eta^2+\chi^2)^2
     \right)\eqno(4.7)$$
both $\eta^2$ and $\chi^4$ couplings. 
Thus, neither of the two fields satisfy $F(S)=0$. Nevertheless, 
we know that there is a flat direction, the circle with radius $v$.
By looking at the lowest order term $\eta^2$ one can read off that
this flat direction locally is $(\eta,\chi)=\varepsilon(0,1)$.
Surely, choosing polar coordinates the angular variable does not appear
in the potential and defines the flat circle. In our case however, 
the local coordinates are given by the primary fields in the SCFT and 
without further knowledge there is no guarantee that these are 
appropriate to capture certain flat directions explicitly. Furthermore,
unlike this model, one generally does not know the superpotential
to all orders, so that extracting definite statements is a difficult
task. In the next section we will exactly be confronted with such
problems, when we try to identify the K\"ahler modulus.
\pano
First, we want to discuss the former sufficient flatness condition in our 
example. In general it is hard to prove in CFT that an infinite set of 
$n-$point functions vanishes unless one is equipped with some selection
rules which a priori disallow certain couplings to be nonzero.
In our case we have a lot of such selection rules related to the special 
enhanced gauge symmetry. On the one hand side both the left and the
right moving $U(1)$ symmetries from each of the N2Vir(k=3) and $U(1)_2$ 
factors have to be preserved. On the other hand side there is the special 
nonabelian $SU(2)$ gauge symmetry, which also constraints the possible 
couplings. In particular, the $SU(2)$ spins have to couple to zero spin in
each correlation function and the relative couplings of members of $SU(2)$
multiplets are determined by Clebsch--Gordan coefficients.
\pano
To make the discussion more transparent we introduce something like
an average, relative charge between the left and right moving sectors of 
the singlets. All the singlets in the Tables 2.2 and 2.3 share the common
feature that in each of the last four N2Vir(k=3) factors five times the 
difference between the left and the right moving $U(1)$ charge
$q_{\rm rel}$ is constant modulo five. For instance, for the field $T'$
$$ \left[0~\matrix{1&1\cr 0&0\cr}\right]
            \left[2~\matrix{-2&0\cr 2&0\cr}\right]
            \left[1~\matrix{-3&-2\cr 1&0\cr}\right]
            \left[1~\matrix{-3&-2\cr 1&0\cr}\right]
            \left[1~\matrix{-3&-2\cr 1&0\cr}\right]
            [1] \eqno(4.8)$$
one has
$$ Q(T')=5\left(\left({3\over5}-1\right)-\left(-{1\over5}\right)\right)
        =5\left(\left({2\over5}\right)-\left(-{2\over5}\right)\right)
        =4\ {\rm mod}\ 5.\eqno(4.9)$$
One can also extend this definition to the other massless states in the
${\bf 16}$ and ${\bf 10}$ representation. In order to make it well--defined
one has to take into account the charge of the $SO(8)$ piece, as well. 
Considering the decomposition of representations of $SO(10)$ in terms of
$SO(8)\times U(1)$ 
$$[16]=[8^1_v]\oplus[8^{-1}_c],\quad\quad[10]=[1^{-2}]\oplus[8^{0}_s]\oplus
 [1^2]\eqno(4.10)$$
the correct definition is
$$Q=5\,q_{\rm rel}-{1\over2}(-1)^\lambda\quad{\rm mod}\ 5.\eqno(4.11)$$
This definition is very similar to the space--time R--charge $S_Q$ 
introduced in [\diskb,\diskc] for the massless states in the 
Landau--Ginzburg model, for it contains information about the twisted 
sector of the $GSO$ projection, in which the massless states occur. 
One obtains for all the massless states in the model Table 4.1 of 
space--time R--charges:
\meno
\cl{\vbox{
\hbox{\vbox{\offinterlineskip
\def\tablespace{height2pt&\omit&&\omit&&\omit&&\omit&&\omit&&\omit&\cr}
\def\tablerule{\tablespace\noalign{\hrule}\tablespace}
\hrule\halign{&\vrule#&\strut\hskip0.2cm\hfil#\hfill\hskip0.2cm\cr
\tablespace
& $S$ && $T'$ && $S'$ && {\bf 16} && {\bf 10} && ${\bf 10'}$ &\cr
\tablerule\tablerule
& $0$ && $4$ && $2$ && $-{1\over 2}$ && $-1$ && $3$ &\cr
\tablespace}\hrule}}
\hbox{\hskip 0.5cm Table 4.1 \hskip 0.5cm R--charges of massless states}}}
\meno
Note that this charges are completely analogous to the R--charges of
the different massless fields in the Landau--Ginzburg model, thus
providing further evidence for the identification of these two models. 
The general form of a superpotential coupling of order $n$ is expressed 
in terms of world sheet operators in the following way:
$$ C^{1,\ldots,n} =\int d^2 z_n\ldots\int d^2 z_1\la V^n_0(z_n,\bz_n)\ldots
   V^4_0(z_4,\bz_4)\,V^3_{-1}(z_3,\bz_3)\,V^2_{-{1\over 2}}(z_2,\bz_2)\,
   V^1_{-{1\over 2}}(z_1,\bz_1)\,\ra.\eqno(4.12) $$
The lower index indicates the ghost picture, in which the vertex operator 
has to be taken. Using the $SL(2,\BC)$ symmetry one can shift in the usual
way three coordinates to $\lbrace 0,1,\infty\rbrace$ and can get rid of 
three integrations by including the correct measure, for instance
$$\int d^2 z_n\ldots\int d^2 z_4|z_1-z_2|^2\,|z_1-z_3|^2\,|z_2-z_3|^2
 \ldots.\eqno(4.13)$$
If the vertex operator for a massless state in a general representation of 
the gauge group has the form 
$$V_{-1}(z,\bz)=e^{-\rho(\bz)}\ \cO _{1}(z,\bz)\ \cF(z)\ \lambda^a(z)\ 
 e^{ikX(z,\bz)}\eqno(4.14)$$
in the (--1) ghost picture, in the $\left(-{1\over 2}\right)$ ghost 
picture it will look like
$$V_{-{1\over2}}(z,\bz)=e^{-{\rho(\bz)\over2}}\ S^\alpha(\bz)\ 
 \Sigma_r\cO_{1}(z,\bz)\ \cF(z)\ \lambda^a(z)\ e^{ikX(z,\bz)}\eqno(4.15)$$
with $S^\alpha (\bz)$ being a four--dimensional spinor and $\Sigma_r$ the 
internal right moving part of the space--time supercharge. In our case, it 
is simply the primary field
$$ \Sigma_r=\left[0~\matrix{0&0\cr 1&1\cr}\right] 
            \left[0~\matrix{0&0\cr 1&1\cr}\right] 
            \left[0~\matrix{0&0\cr 1&1\cr}\right]
            \left[0~\matrix{0&0\cr 1&1\cr}\right]
            \left[0~\matrix{0&0\cr 1&1\cr}\right]\, [0].\eqno(4.16)$$
In the $(0)$ ghost picture one gets
$$ V_{0}(z,\bz)=(G_{r,\rm tot}+k\psi)\ \cO _{1}(z,\bz)\ \cF(z)\
                 \lambda^a(z)\ e^{ikX(z,\bz)},\eqno(4.17)$$
where $G_{r,\rm tot}$ is the total world sheet supercurrent of the $c=9$ 
right moving $N=2$ part:
$$G_{r,\rm tot}=\sum_{i=1}^5\left[0~\matrix{0&0\cr 0&0\cr}\right]\ldots
  \underbrace{\left[0~\matrix{0&0\cr0&2\cr}\right]}_{i\,{\rm th\ factor}}
  \ldots\left[0~\matrix{0&0\cr 0&0\cr}\right]\,[0].\eqno(4.18)$$
One can also attach an R--charge to the two operators $\Sigma_r$ and 
$G_{r,\rm tot}$ appearing in the $-{1\over 2}$ and $0$ ghost picture, 
respectively. Since $Q(\Sigma_r)=-{3\over2}$ and $Q(G_{r,\rm tot})=0$, the 
sum of all $Q$ charges of all internal fields $\cO\,\cF$ must be equal to 
$Q=3$ in order have vanishing R--charge for the entire coupling constant.
Thus, formally a term in the superpotential must have $Q=3$ yielding 
already severe constraints on the possible terms in the superpotential. 
\pano
Surely, Yukawa couplings like $\la 10\ 16\ 16\ra$ or $\la S\ 10\ 10\ra$ 
could take nonzero values but couplings like $\la 10'\ 16\ 16\ra$ or
$\la S'\ 10\ 10\ra$ are forced to be zero by $Q$ conservation. For 
instance, among the couplings of twisted fields only $\la S'\ 10'\ 10'\ra$ 
can also be nonzero. We will come back to such couplings later. Since all 
untwisted singlets have zero R--charge, it follows directly that
$$ F(S)=0,\quad\quad F(S)\,S'=0.\eqno(4.19)$$
However, one must not forget the seven D--flatness conditions for
$SU(2)\times U(1)^4$, so that only 318 of the 325 singlets survive
as moduli of the SCFT. The remaining seven are ``eaten" by the 
super Higgs mechanism. It is not surprising that this is the
same result as already described in [\diskb] for the Landau--Ginzburg 
model.
\pano
At a generic point in the Landau--Ginzburg phase all three--point couplings
are zero, for the four additional fields in the $k=5$ twisted sector do not
occur. However, in our exactly solvable model there exists one coupling 
which can be nonzero by R--charge conservation, namely
$$ \la T'\ S'\ S'\ra. \eqno(4.20)$$ 
By taking into account that every individual left and right moving $U(1)$
charge has to be conserved one finds that actually only the following 
three--point functions have a chance to be nonzero:
$$ \la T'\ S_b^{\prime\, (i)}\ S_c^{\prime\, (j)}\ra, \eqno(4.21)$$
where $i,j=1,0,-1$ are indices of the adjoint representation of $SU(2)$. 
Using the explicit form of the three--point functions of the N2Vir(k=3) 
and $U(1)_2$ models [\zamo], one obtains for this coupling the 
nonvanishing value
$$ C_{T'\,S^{\prime\, (i)}_b\,S^{\prime\, (j)}_c}= 
    \sqrt 3\, \left(\matrix{0 & 1 & 1 \cr 0 & i & j \cr}\right) 
    \kappa^3,\quad\quad \kappa= \sqrt{\Gamma\left({3\over5}\right)^3
    \Gamma\left({1\over5}\right)\over\Gamma\left({2\over5}\right)^3 
    \Gamma\left({4\over5}\right)},\eqno(4.22)$$
where $\left(\matrix{j_1 & j_2 & j_3 \cr m_1 & m_2 & m_3 \cr}\right)$ 
denotes Wigner's $3j$ symbols. Before discussing the resulting obstruction
we move forward and calculate all nonvanishing four--point couplings of 
the gauge singlets. R--charge conservation tells us that there
are only three possible types of such couplings 
$$\la T'\ T'\ S\ S\ra,\quad\quad\la T'\ S'\ S'\ S\ra,\quad\quad
 \la S'\ S'\ S'\ S'\ra.\eqno(4.23)$$
The detailed analysis of all $U(1)$ charges shows that only the following
three couplings of the third type in (4.23) satisfy all selection rules:
$$\la S_a^{\prime}\ S_a^{\prime}\ S_c^{\prime\,i}\ S_c^{\prime\,j}\ra,\quad
\la S_a^{\prime}\ S_b^{\prime\,i}\ S_b^{\prime\,j}\ S_c^{\prime\,k}\ra,
\quad \la S_b^{\prime\,i}\ S_b^{\prime\,j}\ S_b^{\prime\,k}\ 
S_b^{\prime\,l}\ra.\eqno(4.24)$$
As explicitly shown by E.\ Silverstein in [\silvera], the mere fact that a 
four--point function does satisfy all selection rules does not guarantee 
that the superpotential coupling is nonzero. In [\silvera] it was shown 
that the fourth order coupling of certain twisted singlets for the quintic
in $\BC{\rm P}[4]$ miraculously vanishes even though the conformal field 
theoretic four--point function is apparently nonzero. We can follow the 
calculation carried out in [\silvera] for the three fourth order couplings
above (4.24), from which we discuss the first one in more detail. First, 
we want to calculate the four--point function
$$\la V^c_{-1}(z_4,\bz_4)\,V^c_{0}(z_3,\bz_3)\,
V^a_{-{1\over2}}(z_2,\bz_2)\,V^a_{-{1\over2}}(z_1,\bz_1)\,\ra.\eqno(4.25)$$
Since in this case no contact terms can arise, we can
take the zero momentum limit just from the beginning. 
The four vertex operators at zero momentum are 
$$\eqalignno{
 V^a_{-{1\over2}}(z_1,\bz_1)&=e^{-{\rho(\bz_1)\over 2}}\ S^\alpha(\bz_1)
            \left[3~\matrix{-4 &-1\cr 4&1\cr}\right]
            \left[1~\matrix{-1&0\cr 2&1\cr}\right]^2
            \left[0~\matrix{-2&-2\cr 1&1\cr}\right]^2
            [1]\ (z_1,\bz_1), &\cr
 V^a_{-{1\over2}}(z_2,\bz_2)&=e^{-{\rho(\bz_2)\over 2}}\ S^\beta (\bz_2)
            \left[3~\matrix{-4 &-1\cr 4&1\cr}\right]
            \left[0~\matrix{-2&-2\cr 1&1\cr}\right]^2
            \left[1~\matrix{-1&0\cr 2&1\cr}\right]^2
            [1]\ (z_2,\bz_2), &\cr
 V^c_0(z_3,\bz_3)&= 
            \left[1~\matrix{-1 & 0\cr 1&2\cr}\right]
            \left[1~\matrix{-1&0\cr 1&0\cr}\right]^4
            [-2]\ (z_3,\bz_3), &(4.26)\cr
 V^c_{-1}(z_4,\bz_4)&=e^{-{\rho(\bz_4)}}
            \left[1~\matrix{-1 & -2\cr 1&0\cr}\right]
            \left[1~\matrix{-1&0\cr 1&0\cr}\right]^4
            [0]\ (z_4,\bz_4).&\cr }$$
Using N2Vir(k)$={SU(2)_k\over U(1)}\times U(1)$ we split the primary 
fields of N2Vir(k=3) into parafermi\-onic primaries and vertex 
operators of the free boson $\phi$:
$$\left[l~\matrix{q&s\cr\o q&\o s\cr}\right](z,\bz)=
 \phi^l_{q-s,\o{q}-\o{s}}(z,\bz)\ e^{i\alpha_{q,s}\phi(z)}\, 
 e^{i\alpha_{\o{q},\o{s}}\phi(\bz)} \eqno(4.27)$$
with $\alpha_{q,s}={1\over\sqrt{15}}(-q+{5\over2}s)$. 
The correlation functions of the four--dimensional space--time fields and 
the ghost system are quite simple:
$$\eqalignno{
 \la S^\beta(\bz_2)\ S^\alpha(\bz_1)\ra&={\delta_{\alpha\beta}\over
 (\bz_2-\bz_1)^{1\over 2} } &(4.28)\cr\la e^{-\rho (\bz_4)}\ 
 e^{-{\rho (\bz_2)\over 2}}\ e^{-{\rho (\bz_1)\over 2}} \ra &=
 {1\over (\bz_2-\bz_1)^{1\over 4}\, (\bz_4-\bz_1)^{1\over 2}\,
 (\bz_4-\bz_2)^{1\over 2}}.&\cr } $$
Now, by using $SL(2,\BC)$ we set $z_4=0$, $z_2=1$ and $z_1=\infty$ and 
realize that the correlation functions (4.28) and the measure in (4.13) 
are independent of the variable $z_3=:x$. The correlation functions for the
vertex operators in (4.27) and the $U(1)_2$ piece can be expressed in terms
of $x$ as
$$\la\ldots\ra_{U(1)}=|x|^{-{4\over3}}\,|1-x|^{-{4\over3}}.\eqno(4.29)$$
Thus, it only remains to determine five four--point functions for the 
parafermionic piece:
$$\eqalignno{P_1&=\la \phi^0_{0,0}(z_4,\bz_4)\ \phi^0_{0,0}(z_3,\bz_3)\ 
        \phi^1_{-1,-1}(z_2,\bz_2)\ \phi^1_{1,1}(z_1,\bz_1)\ra &\cr
        P_2&=P_3=\la \phi^1_{-1,1}(z_4,\bz_4)\ \phi^0_{0,0}(z_3,\bz_3)\
        \phi^1_{-1,1}(z_2,\bz_2)\ \phi^1_{-1,1}(z_1,\bz_1)\ra &(4.30)\cr 
        P_4&=P_5=\la \phi^0_{0,0}(z_4,\bz_4)\ \phi^1_{-1,1}(z_3,\bz_3)\
        \phi^1_{-1,1}(z_2,\bz_2)\ \phi^1_{-1,1}(z_1,\bz_1)\ra. &\cr}$$
In contrast to the four--point function in [\silvera], here all 
parafermionic amplitudes can be expressed in terms of two-- and 
three--point functions. These are well known [\zamo], so that we arrive 
for the parafermionic correlation function at
$$\prod_{i=1}^5 P_i=\kappa^4\ |x|^{-{4\over15}}\,|1-x|^{-{4\over15}}.
 \eqno(4.31)$$ 
Inserting (4.29,4.31) into the superpotential coupling (4.25), one 
finally obtains
$$\eqalignno{
 \la S_c^{\prime\,(i)}\ S_c^{\prime\,(j)}\ S_a^{\prime}\ S_a^{\prime}\ra&=
    \sqrt 3\,\left(\matrix{0 & 1 & 1 \cr 0 & i & j \cr}\right)\
    \int d^2 x\ \kappa^4\ |x|^{-{8\over 5}}\, |1-x|^{-{8\over 5}} &\cr
    &= \sqrt 3\, \left(\matrix{0 & 1 & 1 \cr 0 & i & j \cr}\right)\
   \kappa^4\ B\left({1\over5},{1\over5},{3\over5}\right) &(4.32)\cr }$$
with
$$ B(a,b,c)=\pi\,{\Gamma(a)\Gamma(b)\Gamma(c)\over
 \Gamma(a+b)\Gamma(b+c)\Gamma(c+a)}.\eqno(4.33) $$
This coupling is finite and nonzero. In the same way, one 
obtains for the second coupling in (4.24)
$$\eqalignno{
 \la S_a^{\prime}\ S_b^{\prime\,(i)}\ S_b^{\prime\,(j)}\ S_c^{\prime\,(k)}
 \ra&=-{3\over\sqrt 2} \left(\matrix{1&1&1\cr-(i+j)&i&j\cr}\right)\,
 \left(\matrix{0 & 1 & 1 \cr 0 & i+j & k \cr}\right)\ 
 \int d^2 x \kappa^4\ |x|^{-{6\over 5}}\, |1-x|^{-{8\over 5}} &\cr
 &= -{3\over\sqrt 2} \left(\matrix{1&1&1\cr-(i+j)&i&j\cr}\right)\,
 \left(\matrix{0 & 1 & 1 \cr 0 & i+j & k \cr}\right)\
 \kappa^4\ B\left({2\over 5}, {1\over 5},{2\over 5}\right). &(4.34)\cr }$$
The calculation of the most complicated third coupling fortunately is 
exactly the same as for the twisted fourth order coupling in (4.24) 
implying that it vanishes after performing the integral over
the complex plane,
$$\la S_b^{\prime}\ S_b^{\prime}\ S_b^{\prime}\ S_b^{\prime}\ra =0.
 \eqno(4.35)$$
From the conformal field theory point of view we have yet no understanding
why this happens, in particular the arguments of [\silverb] suggest 
that the corresponding twisted fields for the quintic are truly moduli, 
so that all couplings involving this field should vanish. 
\meno
\section{5.\ Consequences for the (0,2) moduli space}
\meno
The few nonzero superpotential couplings calculated so far do have already
interesting consequences for the moduli space. However, first we want to
discuss the four singlets $T'$. Since singlets of this kind do not occur at
a general point in the Landau--Ginzburg model, it is tempting to identify 
them with those appearing in the $k=5$ twisted sector for our special 
choice of the polynomials (3.2). Hence, one would expect these fields to 
get a mass, when one generically deforms the complex and bundle structure.
Indeed, even though we cannot calculate them exactly, there exist, for 
instance, sixth order couplings like
$$\la T^{\prime}\ T^{\prime}\ S_h\ S_h\ S_h\ S_h\ra\eqno(5.1)$$
which can create a mass for the singlets $T'$. Since already the 
three--point coupling (4.22) containing the singlet $T'$ is finite, one 
does not expect miraculous cancellations for other couplings.
\pano
Besides the 318 untwisted moduli one expects at least 11 further flat 
directions from the twisted sector. Do the first two orders of the
superpotential allow so many moduli? As the simple Higgs potential
has taught us,
this question is hard to answer, for surely we do not know the entire
superpotential. However, let us mention the following observation:\
Suppose, one can find some untwisted singlets so that giving a 
VEV to the field $S_c^{\prime\, (0) }$ and these untwisted singlets 
satisfies D--flatness. Then the three--point coupling gives masses to
the four fields $T'$ and four of the twisted fields $S'_b$. 
Furthermore, the four--point coupling (4.32) generates mass terms
for the six fields $S'_a$. Thus, including the Higgs effect we are
left with exactly $329$ massless fields, which is the number expected
from the linear $\sigma-$model. As we will show below, the field
$S_c^{\prime\, (0) }$ is not exactly the K\"ahler modulus, but a linear
combination of the fields $S'$. But we are confident that the surviving 
number of $329$ massless states is stable under ``rotation" of 
$S_c^{\prime\, (0)}$ to $R$. The complexified K\"ahler moduli space of 
the linear $\sigma-$model can be sketched like in Figure 1.
\pano
  \centerline{\epsfxsize=13.9cm\epsfbox{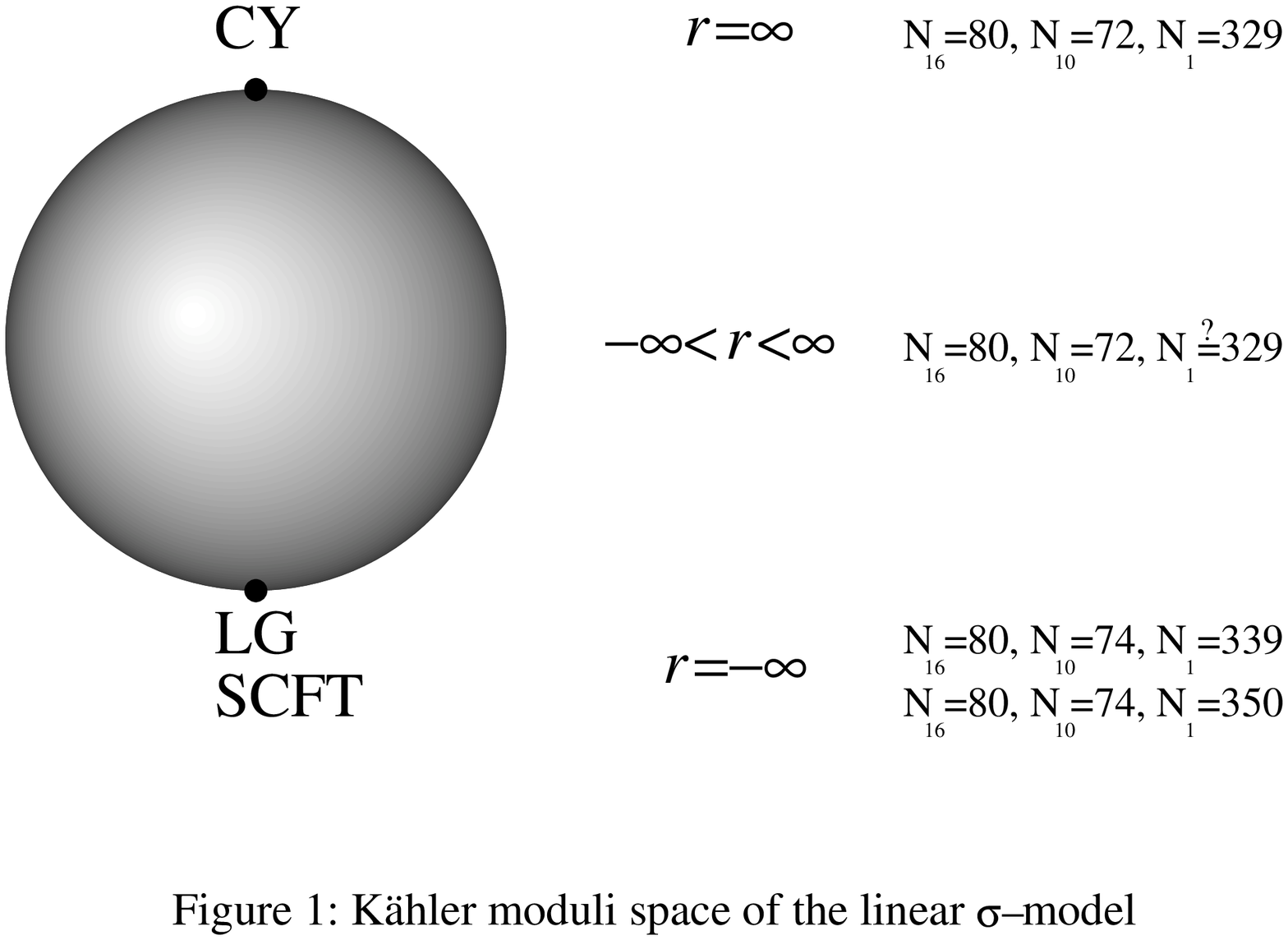}}
\pano
In $(2,2)$ compactifications complex and K\"ahler moduli are related to
the $27$ and $\o{27}$ matter fields by the action of the
left moving supercurrent. Thus, there is an algebraic distinction
among complex, K\"ahler and gauge bundle moduli even for
small radius. In the $(0,2)$ case there exist no left moving supercurrent,
so that a priori there is no way to decide to which class of moduli
a given singlet belongs. In the following, we will show how
to use the calculated couplings to determine the form of the K\"ahler
modulus at least to lowest order. 
\pano
The following properties are expected from a modulus leading away from
the Landau--Ginzburg radius $r=-\infty$:
\pano
\item{a.)} We completely know the scalar potential in the renormalizable
           limit. Thus, in order to determine the local flat direction
           we require D--flatness (4.4) and F--flatness up to cubic 
           couplings.
\pano
\item{b.)} By deforming the radius we expect to obtain the massless 
           spectrum of the large radius CY limit. In particular, the two 
           twisted chiral multiplets in the vector representation of 
           $SO(10)$ should gain a mass. 
\pano
\item{c.)} The special enhanced gauge group $SU(2)\times U(1)^4$ is a pure
           stringy effect and therefore not present in the CY phase.
\pano
\item{d.)} The massless spectrum features the permutation symmetry $S_4$ 
           in the last four tensor factors. We require that the unique 
           K\"ahler modulus $R$ also has this $S_4$ symmetry. 
\pano
\item{e.)} The untwisted $SO(10)$ singlets are given by polynomials of 
           degree four modulo some relations. This is analogous to the
           complex and bundle deformations of the Calabi--Yau manifold. 
           Consequently, we expect $R$ to have contributions only from the
           $S'$ twisted sector.
\meno 
We start with point b.)\ and calculate the Yukawa coupling 
$\la 1\ 10\ 10\ra$ for the two extra ${\bf 10'}$s from the twisted 
sector. They are of the form listed in Table 5.1 in the $[1^{-2}]$ 
sector of $SO(8)$. 
\meno
\cl{\vbox{
\hbox{\vbox{\offinterlineskip
\def\tablespace{height2pt&\omit&&\omit&&\omit&&\omit&\cr}
\def\tablerule{\tablespace\noalign{\hrule}\tablespace}
\hrule\halign{&\vrule#&\strut\hskip0.2cm\hfil#\hfill\hskip0.2cm\cr
\tablespace
&Type&&\hskip4.4cm$\cO_1$\hskip4.3cm$\cF$&&rep.&&deg.&\cr
\tablerule\tablerule
& $G$ && $\left[1~\matrix{-1&0\cr 1&0\cr}\right]
          \left[1~\matrix{-1&0\cr 1&0\cr}\right]
          \left[1~\matrix{-1&0\cr 1&0\cr}\right]
          \left[1~\matrix{-1&0\cr 1&0\cr}\right]
          \left[1~\matrix{-1&0\cr 1&0\cr}\right] [2]$
&& $2$ && $2$ & \cr
\tablespace}\hrule}}
\hbox{\hskip 0.5cm Table 5.1 \hskip 0.5cm Twisted ${\bf 10'}$s}}}
\meno
Now, one can look for Yukawa couplings containing two
twisted ${\bf 10'}$s and one twisted singlet $S'$. The selection
rules allow only one such coupling, which can be
calculated in the usual way: 
$$ \la S^{\prime\, (i)}_c\ G^{(j)}\ G^{(k)}\ra ={1\over 3}
 \left(\matrix{1&{1\over2}&{1\over2}\cr-(j+k)&j&k\cr}\right)\,
 \left(\matrix{0&1&1\cr0&i& j+k\cr}\right)\,\kappa^3.\eqno(5.2)$$
Thus, deforming in the direction of the singlet $S'_c$ gives $G$ a mass
and one is left with the large radius limit for the number of ${\bf 10}$s.
However, since the $SU(2)$ triplet $S'_c$ alone can not break the $SU(2)$ 
gauge group completely, $R$ must contain contributions from other twisted
states. The most general ansatz compatible with b.)$\,-\,$e.) is
$$ R=\sum_{j=-1}^1 \gamma_j\ S_c^{\prime\,(j)} + 
        \sum_{j=-1}^1 \sum_{m=1}^4 \beta_j\ S_{b,m}^{\prime\,(j)} +
        \sum_{n=1}^6 \alpha\ S_{a,n}^{\prime}.\eqno(5.3) $$
Plugging this ansatz into the flatness conditions (4.4) allows up to 
gauge transformations only the following two parameter solution:
$$ R = \gamma\, S_c^{\prime\,(0)}+
     \sum_{m=1}^4 \beta\, \left( \ S_{b,m}^{\prime\, (-1)} +
                  \ S_{b,m}^{\prime\, (+1)}\right) +
     \sum_{n=1}^6 {\gamma\over\sqrt 6}\ S_{a,n}^{\prime}. \eqno(5.4)$$
Using the four--point functions (4.32--4.34), one can show that
$R^4\ne 0$ for all $\gamma,\beta\in\BC,\gamma\ne 0$. Since we know that 
there must exist a flat solution and up to first order the solution is 
highly restricted, we conclude that the nonvanishing four--point coupling 
indicates merely that the conformal fields are an unappropriate basis, in 
which $R$ is curved. The solution (5.4) merely gives the tangent vector at 
the SCFT point along this curve. In order to determine the next order 
corrections to $R$ one also has to take into account the nonflat 
$K(\phi,\phi^*)$. The picture so far obtained for the K\"ahler modulus
is shown in Figure 2.
\pano
  \centerline{\epsfxsize=13.9cm\epsfbox{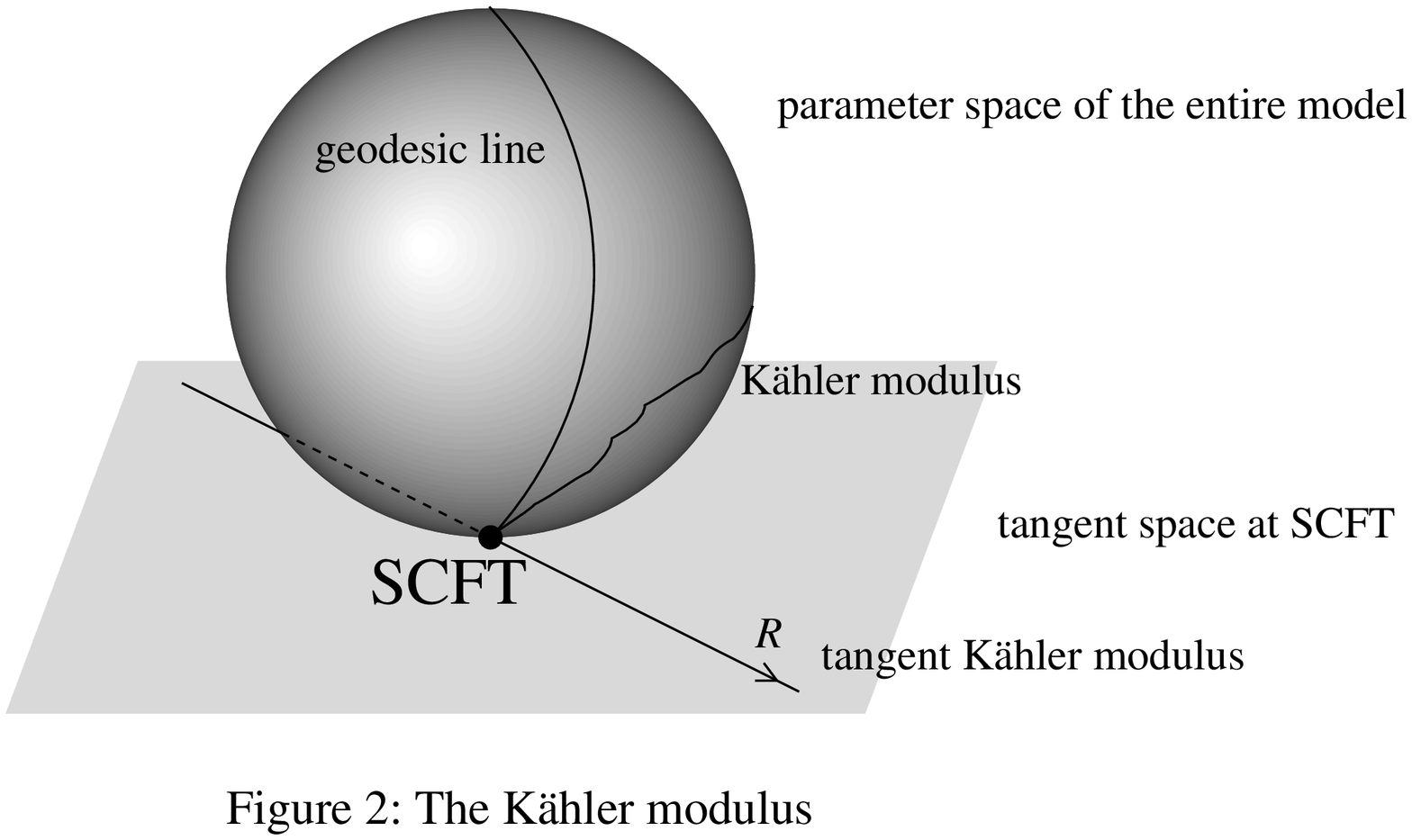}}
\pano
The sphere visualizes the parameter space of the entire model. The
superpotential is a function (or better a section) over this space.
In this parameter space there is a flat direction $R$, of which our
first order calculation only determines the tangent vector at the SCFT
point. If the SCFT would yield appropriate coordinates (like the
polar coordinates in the Mexican hat example), then the flat direction
would be a geodesic line on the sphere. 
\pano
Knowing the singlet associated with the K\"ahler modulus allows one in 
principle to investigate couplings of the form
$$\la R^n\ 10\ 16\ 16\ra,\quad\quad\quad\la R^n\ 1\ 10\ 10\ra.\eqno(5.5)$$
A nonvanishing coupling would detect a radius dependence of the Yukawa 
couplings, which perturbatively was argued to be absent [\disgreb,\greene].
Unfortunately, the selection rules do not forbid couplings
of the form (5.5) and their exact calculation is hard to come by. 
\pano
Since at the Landau--Ginzburg point there generically occur ten more
$SO(10)$ singlets than in the Calabi--Yau phase, it is tempting
to speculate about new flat directions of the superpotential
leading perhaps to completely new phases of $(0,2)$ models or
even to a different linear $\sigma-$model [\diskd]. 
However, the data achieved so far neither rule out nor seem to prove such a
possibility. We have tried to find a flat direction with a contribution
from the $T'$ field and only other twisted singlets. Since $T'$
only appears at the very special SCFT point such a flat direction
could not be part of the linear $\sigma-$model moduli space. 
However, for such an ansatz no solution to first order exist. 
If one also allows contributions from the untwisted singlets
to first order there exist plenty of solutions, as for instance
$$N=\alpha\left(\ \sum_{m=1}^4\left(S_{e,m}^{({1\over2})}-
    S_{e,m}^{(-{1\over2})}\right)+\sum_{n=1}^4{\sqrt 2}\left(
    S_{g,n}^{({1\over2})}+S_{g,n}^{(-{1\over2})}\right)+
    \sum_{p=1}^4T'_p\right).\eqno(5.6)$$
Finally, we want to mention a coincidence, which might perhaps lead to a 
better understanding of the $(0,2)$ moduli space. The number of scalars in
the vector representation of $SO(10)$ is $N_{10}=74$, which is the same as
the sum of complex moduli ($b_{21}=73$) and K\"ahler moduli ($b_{11}=1$) of
the underlying Calabi--Yau manifold.
\meno
\section{6.\ Summary}
\meno
In this paper we have provided further convincing arguments for the
identification of an exactly solvable $(0,2)$ string 
model with a special point in the Landau--Ginzburg phase of a $(0,2)$ 
linear $\sigma-$model. Then, for the first time, we calculated exactly all
three-- and four--point couplings in the space--time superpotential 
yielding obstructions against the deformation in all 350 directions
simultaneously. Similarly to the Landau--Ginzburg analysis the nontwisted
moduli could be derived simply by selection rules. Furthermore, to lowest
order we have up to two parameters identified the singlet corresponding to
the K\"ahler modulus. Unfortunately, the available data did not allow us to
find a unique solution for the K\"ahler modulus showing again the 
difficulty in making $(0,2)$ models technically as well treatable as 
$(2,2)$ models. Since we did not know the entire superpotential, we could 
only speculate about the possibility of further flat directions leading 
perhaps to another linear $\sigma-$model.
\meno
{\bf Acknowledgements}
\smno
It is a pleasure to thank L.\ Dolan, S.\ Kachru, W.\ Nahm, 
R.\ Schimmrigk, E.\ Silverstein and E.\ Witten for discussion. This work
is supported by U.S.\ DOE grant No.\ DE--FG05--85ER--40219.
\meno
\section{References}
\meno
\bibitem{\bag} J.A.\ Bagger,
{\it Coupling the gauge invariant supersymmetric nonlinear sigma--model
to supergravity}, Nucl.\ Phys.\ {\bf B211} (1983) 302
\bibitem{\banks} T.\ Banks, L.J.\ Dixon, D.\ Friedan and E.\ Martinec,
{\it Phenomenology and conformal field theory, or Can string theory predict
the weak mixing angle}?, Nucl.\ Phys.\ {\bf B299} (1988) 613 
\bibitem{\berg} P.\ Berglund, C.V.\ Johnson, S.\ Kachru and P.\ Zaugg,
{\it Heterotic Coset Models and $(0,2)$ String Vacua}, 
Nucl.\ Phys.\ {\bf B460} (1996) 252, hep--th/9509170
\bibitem{\selfa} R.\ Blumenhagen and A.\ Wi{\ss}kirchen, {\it Exactly 
solvable $(0,2)$ supersymmetric string vacua with GUT gauge groups},
Nucl.\ Phys.\ {\bf B454} (1995) 561, hep--th/9506104 
\bibitem{\selfb} R.\ Blumenhagen, A.\ Wi{\ss}kirchen and R.\ Schimmrigk,
{\it The $(0,2)$ exactly solvable structure of chiral rings, 
Landau--Ginzburg theories and Calabi--Yau manifolds}, 
Nucl.\ Phys.\ {\bf B461} (1996) 460, hep--th/9510055
\bibitem{\selfc} R.\ Blumenhagen and A.\ Wi{\ss}kirchen,{\it Exactly 
solvable points in the moduli space of heterotic $N=2$ strings}, 
preprint IFP--606--UNC, BONN--TH--96--01, hep--th/9601050
\bibitem{\canda} P.\ Candelas, G.T.\ Horowitz, A.\ Strominger and
E.\ Witten, {\it Vacuum configurations for superstrings},
Nucl.\ Phys.\ {\bf B258} (1985) 46
\bibitem{\candb} P.\ Candelas and X.\ De la Ossa,
{\it Moduli space of Calabi--Yau manifolds},
Nucl.\ Phys.\ {\bf B355} (1991) 455
\bibitem{\crem} E.\ Cremmer, S.\ Ferrara, L.\ Girardello, B.\ Julia, 
J.\ Scherk and P.\ van Nieuwenhuizen,
{\it Spontaneous symmetry breaking and Higgs effect in supergravity
without cosmological constant}, Nucl.\ Phys.\ {\bf B147} (1979) 105
\bibitem{\dine} M.\ Dine, N.\ Seiberg, X.G.\ Wen and E.\ Witten,
{\it Nonperturbative effects on the string world sheet I+II},
Nucl.\ Phys.\ {\bf B278} (1986) 769, {\sl ibid.\ }{\bf B289} (1987) 319
\bibitem{\disgrea} J.\ Distler and B.\ Greene,
{\it Aspects of $(2,0)$ string compactifications},
Nucl.\ Phys.\ {\bf B304} (1988) 1
\bibitem{\disgreb} J.\ Distler and B.\ Greene,
{\it Some exact results on the superpotential from Calabi--Yau
compactifications}, Nucl.\ Phys.\ {\bf B309} (1988) 295
\bibitem{\diska} J.\ Distler and S.\ Kachru, {\it $(0,2)$ Landau--Ginzburg
theory}, Nucl.\ Phys.\ {\bf B413} (1994) 213, hep--th/9309110
\bibitem{\diskb} J.\ Distler and S.\ Kachru,
{\it Singlet couplings and $(0,2)$ models},
Nucl.\ Phys.\ {\bf B430} (1994) 13, hep--th/9406090
\bibitem{\diskc} J.\ Distler and S.\ Kachru,
{\it Quantum symmetries and stringy instantons},
Phys.\ Lett.\ {\bf B336} (1994) 368, hep--th/9406091
\bibitem{\diskd} J.\ Distler and S.\ Kachru, {\it Duality of $(0,2)$ 
string vacua}, Nucl.\ Phys.\ {\bf B442} (1995) 64, hep--th/9501111
\bibitem{\gepe} D.\ Gepner, {\it Yukawa couplings for
Calabi--Yau string compactifications},
Nucl.\ Phys.\ {\bf B311} (1988) 191
\bibitem{\greene} B.R.\ Greene, {\it Superconformal compactifications
in weighted projective space},
Commun.\ Math.\ Phys.\ {\bf 130} (1990) 335
\bibitem{\intri} K.\ Intriligator, {\it Bonus symmetry in conformal
field theory}, Nucl.\ Phys.\ {\bf B332} (1990) 541
\bibitem{\kawit} S.\ Kachru and E.\ Witten, {\it Computing the complete 
massless spectrum of a Landau--Ginzburg orbifold},
Nucl.\ Phys.\ {\bf B407} (1993) 637, hep--th/9307038
\bibitem{\kawai} T.\ Kawai and K.\ Mohri,
{\it Geometry of $(0,2)$ Landau--Ginzburg orbifolds}, 
Nucl.\ Phys.\ {\bf B425} (1994) 191, hep--th/9402148
\bibitem{\sche} A.N.\ Schellekens and S.\ Yankielowicz, {\it Extended 
chiral algebras and modular invariant partition functions}, Nucl.\ Phys.\
{\bf B327} (1989) 673
\bibitem{\schz} A.N.\ Schellekens and S.\ Yankielowicz, {\it Modular 
invariants from simple currents. An explicit proof}, Phys.\ Lett.\ 
{\bf B227} (1989) 387
\bibitem{\schd} A.N.\ Schellekens and S.\ Yankielowicz, {\it New modular 
invariants for $N=2$ tensor products and four--dimensional strings}, 
Nucl.\ Phys.\ {\bf B330} (1990) 103
\bibitem{\schv} A.N.\ Schellekens and S.\ Yankielowicz, {\it Simple 
currents, modular invariants and fixed points}, Int.\ J.\ Mod.\ Phys.\ 
{\bf A5} (1990) 2903
\bibitem{\silvera} E.\ Silverstein, {\it Miracle at the Gepner point},
Phys.Lett.{\bf B352}(1995)69,\ hep--th/9503150 
\bibitem{\silverb} E.\ Silverstein and E.\ Witten,
{\it Criteria for conformal invariance of $\,(0,2)$ models},
Nucl.\ Phys.\ {\bf B444} (1995) 161, hep--th/9503212
\bibitem{\wittena} E.\ Witten,
{\it New issues in manifolds of $SU(3)$ holonomy},
Nucl.\ Phys.\ {\bf B268} (1986) 79
\bibitem{\wittenb} E.\ Witten,
{\it Phases of $N=2$ theories in two dimensions},
Nucl.\ Phys.\ {\bf B403} (1993) 159, hep--th/9301042
\bibitem{\zamo} A.B.\ Zamolodchikov and V.A.\ Fateev,
{\it Parafermionic currents in two--dimensional conformal quantum field
theory and selfdual critical points in $Z(N)$ symmetric statistical
systems}, Sov.\ Phys.\ JETP {\bf 62} (1985) 215
\vfill
\end